\documentclass[preprint,authoryear,12pt]{elsarticle}
\usepackage{subfigure}
\usepackage{amssymb}
\journal{Advances in Space Research}

\begin{document}

\begin{frontmatter}

\title{Quasi 9  and  30\,--\,40 days periodicities in the solar
 differential rotation}

\author{J. Javaraiah}

\address{Indian Institute of Astrophysics, Bangalore - 560034, India \\
Tel: 91 (80) 25530672,  Fax: 91 (80) 25534043}

\ead{jj@iiap.res.in}

\begin{abstract}
 Using the daily Mt. Wilson Doppler velocity data
during 1986\,--\,1994 (solar cycle~22), we studied the short-term
variations of the order of a few days to a month timescales
in the solar differential rotation. We represent the
differential rotation in the form:
$\omega(\lambda) = \bar A + \bar B (5 \sin^2 \lambda - 1) + \bar C (21 \sin^4 \lambda - 
14 \sin^2 \lambda +1)$,
 using a set of Gegenbauer polynomials, where $\omega(\lambda)$ is the 
angular velocity at latitude $\lambda$.
The coefficients $\bar A$, $\bar B$, and $\bar C$ are free of
crosstalk.
 We found that   $\approx$ 9-day periodicity is statistically highly 
significant  
in  the variations of $\bar C$ at the maximum of
solar cycle~22. A similar periodicity is
found  in the variations of $\bar B$
during the descending phase of the cycle~22 with significant on
 $\ge$ 99.9\% confidence level.
 At this cycle maximum, a 
 30\,--\,40 day periodicity is found to be dominant among the variations in 
$\bar B$, and this periodicity is found  in $\bar A$ during almost throughout
 the period 1986\,--\,1994.
The
$\approx$ 9-day periodicity
in the variation of the differential rotation approximately matches with the
known quasi 10-day periodicity in the total solar irradiance (TSI)
 variability. Hence,
we speculate that there exists a relationship between the differential
 rotation and TSI variability. We suggest that the 9\,--\,10 day
periodicities
 of the differential
rotation and TSI  have a relationship with
 the production and the emergence rates of
the large-scale solar magnetic flux.
\end{abstract}

\begin{keyword}
solar rotation \sep solar magnetic field \sep solar activity
\end{keyword}

\end{frontmatter}

\parindent=0.5 cm

\newpage
\section{Introduction}
Besides the well known $\approx$ 11-year solar cycle,
solar activity varies on many short-time scales (a few minutes to a few years)
 and many long time scales (a few decades to the Sun's evolution
time scale). It has been generally
 accepted that interactions of the Sun's differential rotation (DR)
and the solar magnetic
 field play a basic role in generation of solar activity.
However, the exact role of the DR in the variations of the solar 
activity is not known and the mechanism of solar cycle is not yet
 fully understood. Therefore,  studies on the temporal variations of the DR are 
important for understanding
 the role of the  DR in the  solar variability. 
So far many attempts have 
been made to determine  the varitions in the DR and detected several 
periodicities in  the DR
by using the different
data and different techniques
\citep[$e.g.$,][]{Hlb80,lh82,Gh84,bvw86,khh93,jg95,jk99,gupta99,howe00a,jj03,jj05,jbu05,brw06,ju06,tlat07,jubb09,cvi10}. A higher rotation rate during the cycle minimum than
during the maximum is found in many studies~\citep[$e.g.$,][]{brw06}. 

Variation of the DR with  the 11-year solar cycle is well established now. 
The 11-year torsional oscillations were
 discovered by~\cite{Hlb80},
using Mt. Wilson Doppler velocity measurements have been
 confirmed using different data sets and methods~\citep[see][]{jg02}.
The torsional oscillations consist of alternating bands of faster and slower 
than average rotation moving from high latitudes toward the equator. The faster than average bands start at mid-latitudes of about $45^\circ$ during the 
minimum phase of the solar cycle and move equatorward with the rising phase of
 the new cycle. The bands extend over about $10^\circ$ in latitude. 
The faster than average rotation band is located on the equatorward side 
of the magnetic activity belt and a slower than average rotation band is 
located on its poleward side. Thus, it is thought that the torsional 
oscillations are associated with magnetic shear. At high latitudes, 
the rotation changes from slower to faster than average at (or just after) 
the solar maximum, which might be related to the polarity reversal of polar 
magnetic fields, and remains faster than average to the next solar minimum 
without clear migratory character. 
Helioseismic observations show that the torsional
oscillations are not just
a superficial phenomenon but they extend to at least the upper
third of the solar convection zone~\citep{howe00b,abc08}.
 Variations  on the few other time scales
in the coefficients of solar differential
rotation~\citep{jg95,jg97,jk99,jbu05}
and the residual rotation~\citep{brw06},
including
one which is approximately equal to the Gleissberg cycle,
 have been found using   sunspot group
data.

It is known that 
the total solar irradiance (TSI)  varies by about 0.1\% over the
 solar cycle.
A number of statistical
 models of TSI variability have been constructed on the basis of
inhomogeneities of surface magnetic field. These models helped to identify
the surface magnetic structures (sunspots and faculae) responsible for
  variations
in  TSI and  also provided widely accepted theoretical explanations.
 However, the exact mechanism behind the  TSI variation 
  is not known~\citep{f92,kuhn99,skw05,dom09}. 

\citet{kuhn88}
 found that a significant part of the solar cycle variation in  TSI is a
 result of the temporal changes in the latitude-dependent surface
temperature of the Sun. These authors speculated  that 
the DR (turbulent Reynolds
tresses and their perturbations) has a role in  TSI variability. 
 Besides the $\approx$ 11-year periodicity, 
TSI also seems to vary on
timescales which are  much shorter than 11-year.
For example,  a $\approx$ 10-day periodicity is found to be prominent 
in the short variations of TSI~\citep{f92,nik98} 
and the existence of a $\approx$ 30-day
 periodicity in the proxies of TSI is  reported~\cite[$e.g.$,][]{pap92}.
 However, the amplitudes of the short variations of  TSI 
 depend on the  phase of a solar cycle~\citep[$e.g.$,][]{pap92}. 
 In the present analysis
 we have attempted to find  
the short periodicities in the coefficients of  the DR determined from the
 Mt. Wilson velocity data during 1986\,--\,1994, because which may 
  help for understanding the TSI variability.

In the next section we describe the methodology and the data analysis.
 In Section~3 we present the results, and in Section~4 we present  
 conclusions and briefly discuss them. 

\section{Methodology and data analysis}
The Sun's DR can be determined from full disc velocity data using
the traditional polynomial expansion:
$$\omega(\lambda) = A + B \sin^2\lambda + C \sin^4\lambda, \eqno(1)$$
\noindent where $\omega(\lambda)$ is the solar rotation rate at latitude
$\lambda$, the parameter $A$ represents the equatorial  rotation
 rate, $B$ and $C$ measure the latitude gradient of the rotation rate
with $B$ representing mainly low latitudes and $C$ representing largely
 higher latitudes.

As pointed out by several authors~\citep[see][]{snod84}, 
 due to the non-orthogonality of the fit functions, the coefficients $A$, $B$,
 and $C$ have a crosstalk which affects their temporal behavior.
\citet{sh85}  used the so-called Gegenbauer Polynomials~\citep{mf53} as a
 set of disk-orthogonal fit functions with $T^1_0 (\sin\lambda) = 1$,
 $T^1_2 (\sin\lambda) = 5 \sin^2\lambda - 1$, and
$T^1_4 (\sin\lambda) = 21 \sin^4\lambda - 14 \sin^2 \lambda + 1$, which leads
 to the following expansion:
$$\omega(\lambda) = {\bar A} + {\bar B} (5 \sin^2\lambda - 1) + 
{\bar C}  (21 \sin^4\lambda - 14 \sin^2\lambda  + 1). \eqno(2)$$
\noindent The coefficients $\bar A$, $\bar B$, and $\bar C$ are free of crosstalk,
 $\bar A$ represents the `rigid body' (or `mean') component in the rotation, $\bar B$ and $\bar C$ are the components of the DR. If the polynomial expansion is terminated at $\bar C$ (or C), the coefficients, $\bar A$, $\bar B$, and $\bar C$ are related to the  A, B, and C coefficients as follows:
$${\bar A} = A + (1/5) B + (3/35) C;  \quad {\bar B} = (1/5)B + (2/15)C; \quad
 {\bar C} = (1/21) C. \eqno(3)$$
\noindent In this case, the temporal variation of $\bar C$ is qualitatively 
identical to that of C (Note: in Equation~(6) of \cite{su90} 
1/21 is erroneously typed as 2/21).

Earlier,
 using the  daily values of the $A$, $B$,
 $C$ coefficients (cf., Equation~(1)) derived from the Mt. Wilson velocity data
during the  period 1967\,--\,1994, 
 \citet{jk99} determined nearly one-year and more than one year  
  periodicities in the DR.
 In that early paper the authors
 had focused mainly on the data obtained during
 the period 1982\,--\,1994,  after a major change in the Mt. Wilson spectrograph
modification which reduced the instrumental noise~\citep[see][]{Hlb83}.
Recently, \citet{jubb09}  found the existence of a $\approx$ 1.4-year 
periodicity  in  $A$ determined from the 
Mt. Wilson velocity
data during 1986\,--\,1995,  but they did not find any significant variation
 in $A$  determined from the data  
during 1996\,--2007, which are measured from the
more stable Mt. Wilson spectrograph instrumentation.
However, the 1.3-year periodicity
 in the variation of the low and the middle latitudes' rotation rate 
 near the base of the convection zone, that was detected  by \cite{howe00a} 
from helioseismic measurements during the period 1995\,--\,1999, 
 is also found to be not persisting after 2001 \citep{howe07}.
 In fact, in several solar activity phenomena a 1.3-year periodicity
 has been found to be
 dominant during the cycle~22 and  weak or absent
in the later period~\citep[see][]{os07}.

 The  time series of $A$, $B$, and $C$
 during 1986\,--\,1994 have many data gaps that very in size, 1\,--\,18 days. 
Since here we have 
attempted to determine the periodicities of
 the order of a five days to a month only, hence
it is necessary to analyse the daily data which have no gaps or have 
 only a few 1\,--\,2 day gaps.
We find that the data in the following time (rotation number) intervals
 contain   only a few (two or three) 1\,--\,2 days
gaps:
 7772\,--\,7934, 8197\,--\,8297, 8559\,--\,8726, 8902\,--\,9023, 
9280\,--\,9376, 9583\,--\,9720,
10056\,--\,10161, 10375\,--\,10514 and 10731\,--\,10893. 
 The
 size of each of these intervals is sufficient to determine
 the short periodicities of  a few days to a month 
time. (In the data during a period which is not included in any one of these
 intervals there are large number of  gaps
 of different sizes. Therefore, they are unfit to use for the present purpose.)
Incidentally, each one of these intervals belongs to one of the 
 years 
1986\,--\,1994, in chronological order. That is,  the years 1986,  1987,  
....., 1994  contain 7772\,--\,7934 (first interval),  
8197\,--\,8297 (second interval), ..., 10731\,--\,10893 (last interval), 
respectively. By using Equation~(3), we
converted the $A$, $B$, and $C$ coefficients to the corresponding 
coefficients $\bar A$, $\bar B$, and $\bar C$, respectively. 
We corrected the  time series of each of these coefficients by removing 
 the very large-spikes,  i.e., we have removed the values which are
 $>2.5 \sigma$ level (where $\sigma$ is the standard deviation of the 
original time series).  
 We filled the gaps in each of  these time-series  from the values 
obtained by liner-interpolations.
 Figures~1 and 2 show  these corrected time series and also the corresponding 
 uncorrected time-series. 
As can be seen in these figures, there are
variations of the order of few days to a month time scales   in 
$\bar A$, $\bar B$ and $\bar C$  
during many of the aforementioned 
time intervals.  
We computed fast Fourier transform (FFT) power spectra 
 of $\bar A$, $\bar B$, 
and $\bar C$. The results are presented in the next section.

\section{Results}
  Figures~3\,--\,5 show the FFT power spectra of 
 $\bar A$, $\bar B$ 
and $\bar C$ derived from the corrected  data (solid curves in Figures~1\,--\,2)
 during 
  each of the
 reliable time intervals found above.
Before computing the FFT,  the long-term trend of the order of the 
length  
of the time series was removed by subtracting the corresponding 
 leaner-model of the series and  a cosine bell function was 
applied to the first and the last 10\% of the time 
series~\citep[see][]{bra71}.   
 In Table~1 we have given the
periodicities in the variations of 
 $\bar A$, $\bar B$ and $\bar C$,
whose levels of significance are  
  $\ge$ 2$\sigma$ in the respective FFT power spectra. 
The results in  this table (and Figures~3\,--\,5)    
  suggest that at the solar cycle maximum 
($i.e.$, in the interval 8902--9023 during  the year 1989),
 a  $\approx$9-day periodicity is very pronounced 
in the variations of $\bar C$. In the same time  this   periodicity is 
insignificant/absent
in the variations of both $\bar A$ and $\bar B$, and a 30\,--\,40 day
 periodicity is
dominant in the variations of $\bar B$. During the decay phase of the cycle~22
(years 1991\,--\,1993), there is $\sim$ 9-day periodicity  
 in the variations of
$\bar B$ with $>$ 99\% confidence level. There is a suggestion on the existence 
of the 30\,--\,40 day periodicity 
 in $\bar A$ during almost throughout the period 1986\,--\,1994.

\section{Conclusions and discussion}
By analyzing the  Mt. Wilson Doppler velocity data
during the period 1986\,--\,1994, we find:
\begin{enumerate}
\item During the maximum phase of the solar cycle~22 there
were  highly statistically significant   $\approx$ 9-day  and
 30\,--\,40 day  periodicities in
the DR coefficients  $\bar C$ and $\bar B$, respectively.
\item During the descending phase
of  cycle~22 the $\approx$ 9-day periodicity was stronger
 in $\bar B$ than in $\bar C$, the 30\,--\,40 day periodicity was weak in both
 $\bar B$ and $\bar C$.
\item Near the minimum of cycle~22 the
30\,--\,40 day
periodicity was present  and the $\approx$ 9-day periodicity was 
insignificant/absent
  in $\bar C$.
\item In $\bar A$ the 9-day  periodicity was absent during  
  any phase of
the solar cycle~22, whereas a 30\,--\,40 day periodicity was present  
in almost throughout this cycle.
\end{enumerate}

It is difficult to detect variations in the equatorial rotation rate 
determined from the Doppler velocity measurements
 because several observational
and instrumental effects can produce spurious peaks with
similar periodicities~\citep[$e.g.$,][]{snod83,lw89,su90,sten90,ub96,ws00}.
Variations in the solar rotation on shorter time scales are 
more difficult to detect because of the relatively low amplitudes and 
intermittent nature of them. 

Since 1986 the quality of the Mt. Wilson velocity data is high enough to study
 substantially short variations in the derived velocity values
\citep{snod92,ulr98}, hence,  the dominant $\approx$~9 day and the
30\,--\,40 day
periodicities in $\bar B$ and $\bar C$, found above, may be  really exist in
the  DR. However,
 the amplitudes of these
periodicities in the DR  depend on the phases of
the solar cycle.

A $\approx$ 10-day periodicity seems to  prominently present 
 in the
 variation of the TSI~\citep[see][]{f92,nik98}. 
\cite{pap92} found a $\approx$ 30-day
 periodicity in the proxies of TSI. 
Recently, \cite{shap11} found  13- and 27-day periodicities 
in the  spectral solar irradiance. 
The $\approx$~9-day and 30\,--\,40 day periodicities of
 $\bar B$ and $\bar C$ 
 approximately match with that known $\approx$ 10-day and $\approx$ 
30-day periodicities in
 the TSI variability. 

The existence of
the common periodicities in  the DR and the TSI
  suggests that the latter may be related to the former as
 pointed by~\citet{kuhn88}. Hence, the $\approx$~9-day and 30\,--\,40
 day periodicities  in the  DR, which are detected here, 
 may be of particular importance for understanding
 the TSI
variability.

 It may be
also necessary to note here that the $\approx$~9 
day$^{-1}$ frequency in the DR  may be  the third harmonic of the Sun's 
rotation frequency. The
$\approx$~9-day and $\approx$~30-day periodicities seem to
be predominantly present 
in the solar coronal holes, solar wind, auroral electron power,
 and geomagnetic parameters~\citep{tvv07,emery09}.
Also, helioseismic studies  suggest that global solar oscillation
frequency changes on time scale as short as
 nine days~\citep{trip07}.

\citet{Hlb81} analysed Mt. Wilson magnetograph data during the period
 1967 to mid-1980 and found that the rate at which the
magnetic flux appears on the
 Sun is sufficient to create all the flux that is seen at the solar surface
 within a period of about 10 days. The magnetic structures of the sunspot
groups may rise from near the bottom of the convention zone to 
the surface in about 10 days~\citep{jg97}.
 Thus, the 9\,--\,10 day periodicity
of the DR and the  TSI seems to be related to
 the production and emergence rate  of the large-scale solar magnetic flux.

\vspace{0.5cm}
\noindent{\large \bf Acknowledgments}

\vspace{0.3cm}
 {The author  thanks  Dr. R. F. Howard for kindly providing the data and 
 the  anonymous referees for
 the useful comments and suggestions.
 The author is grateful to  Dr. Rudi Komm for the helpful discussion  
on the preliminary results, at NSO Sac Peak in March  2002, 
  and to Dr. John Leibacher for the financial support.} 

\clearpage

\newpage
\begin{figure}
\begin{center}
{\subfigure{
\includegraphics[width=5cm]{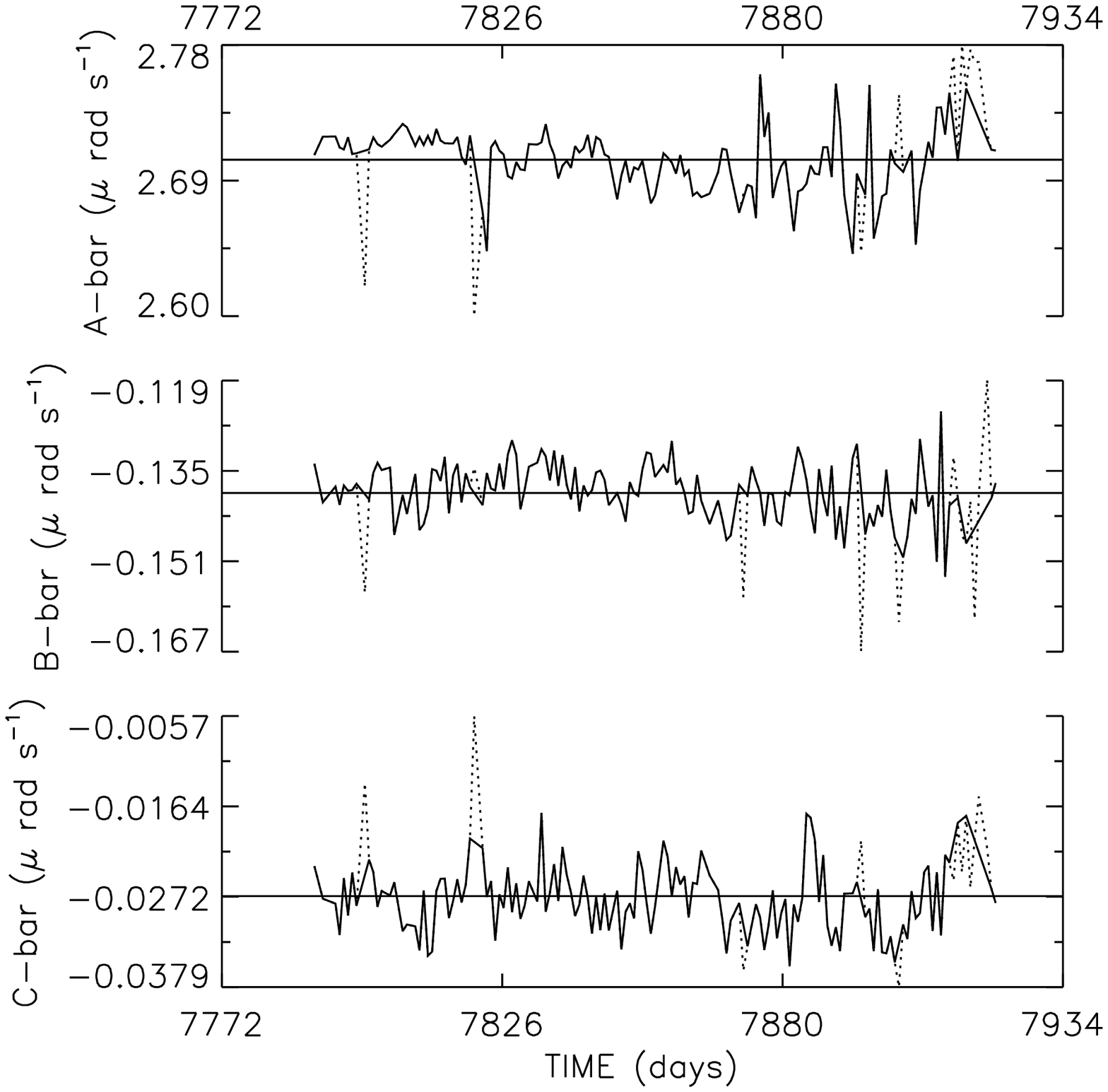}}
\hspace{1.5cm}
\subfigure{
\includegraphics[width=5cm]{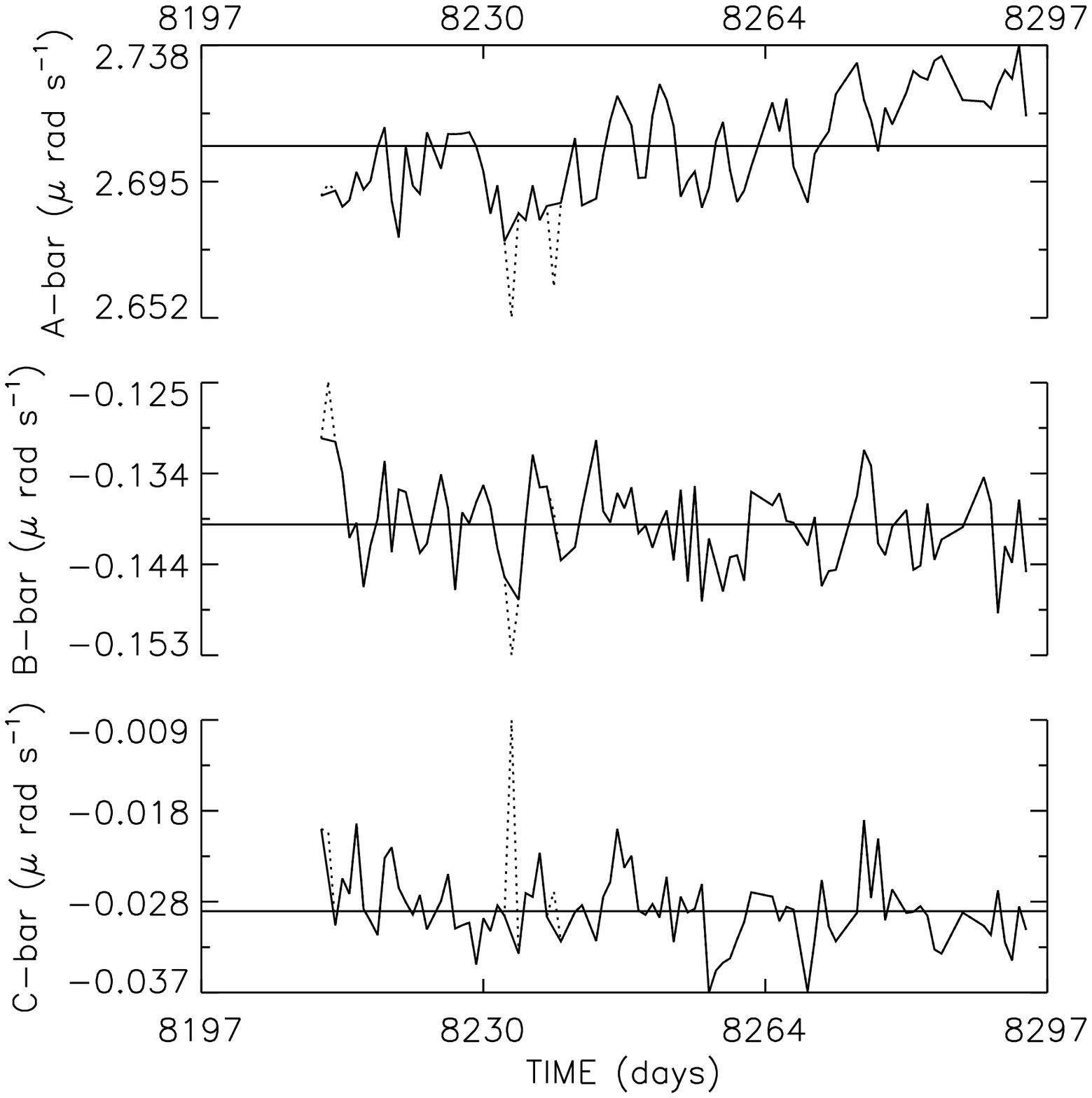}}

\subfigure{
\includegraphics[width=5cm]{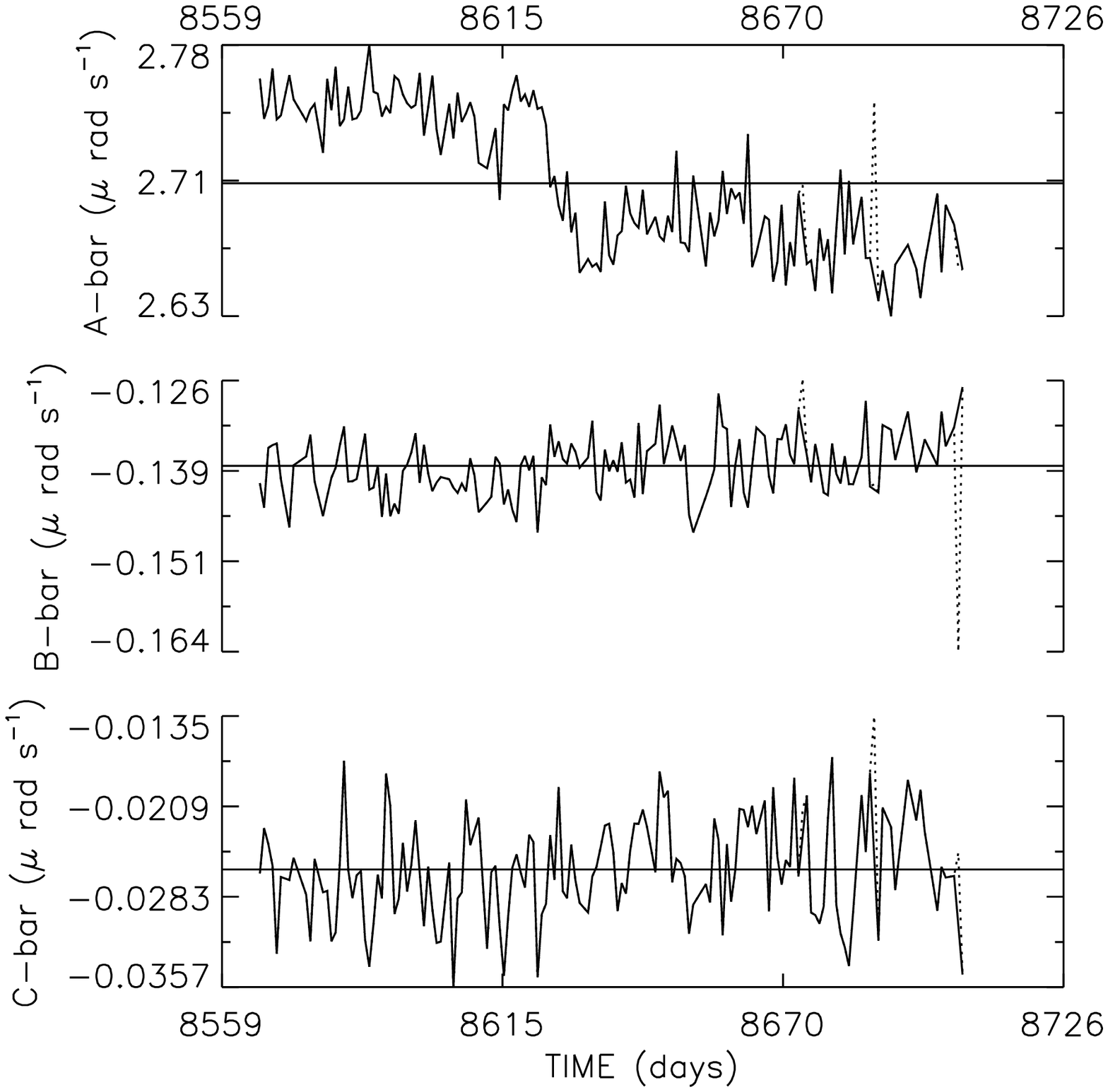}}}
\hspace{1.5cm}
{\subfigure{
\includegraphics[width=5cm]{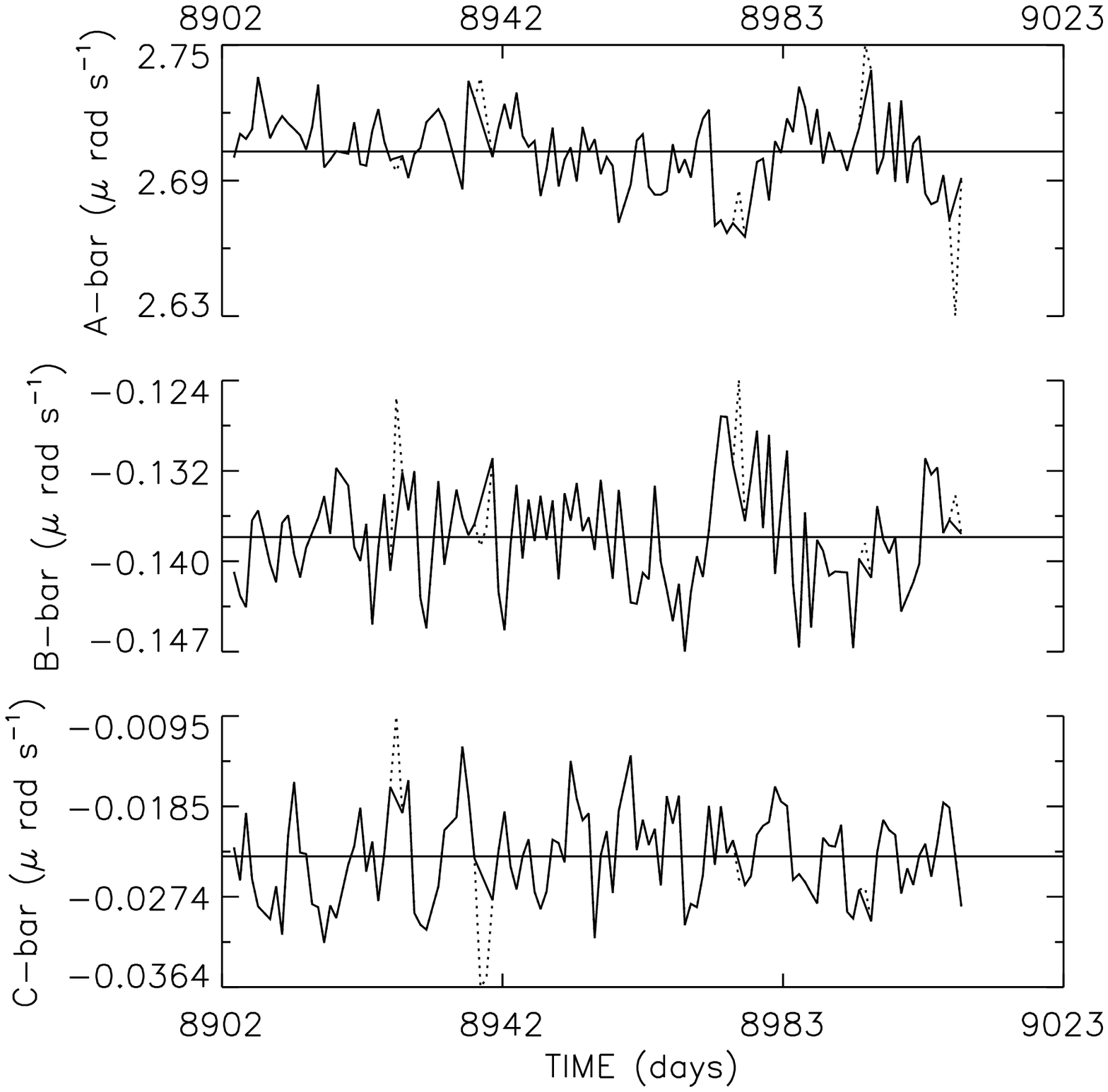}}

\subfigure{
\includegraphics[width=5cm]{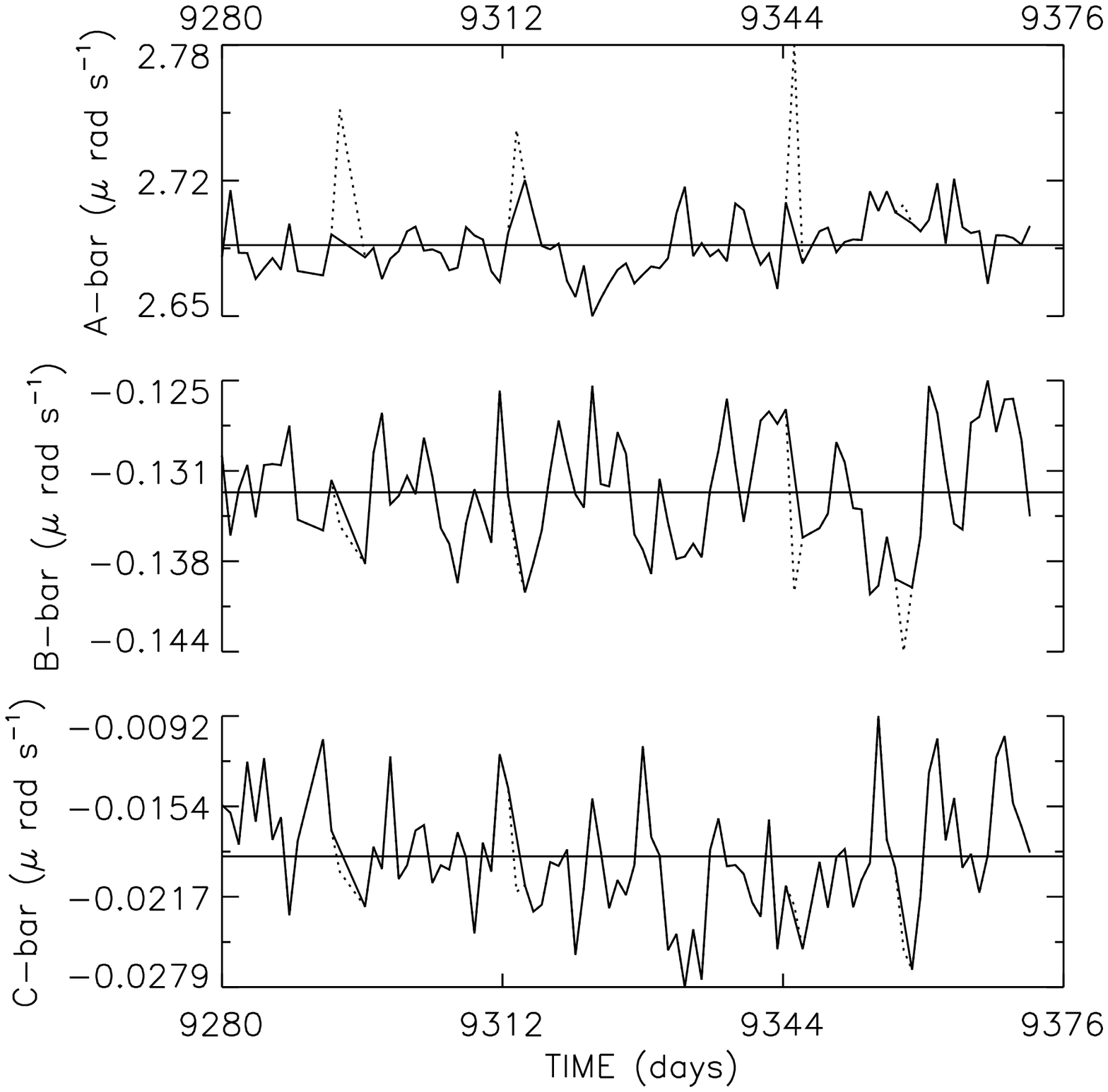}}
\hspace{1.5cm}
\subfigure{
\includegraphics[width=5cm]{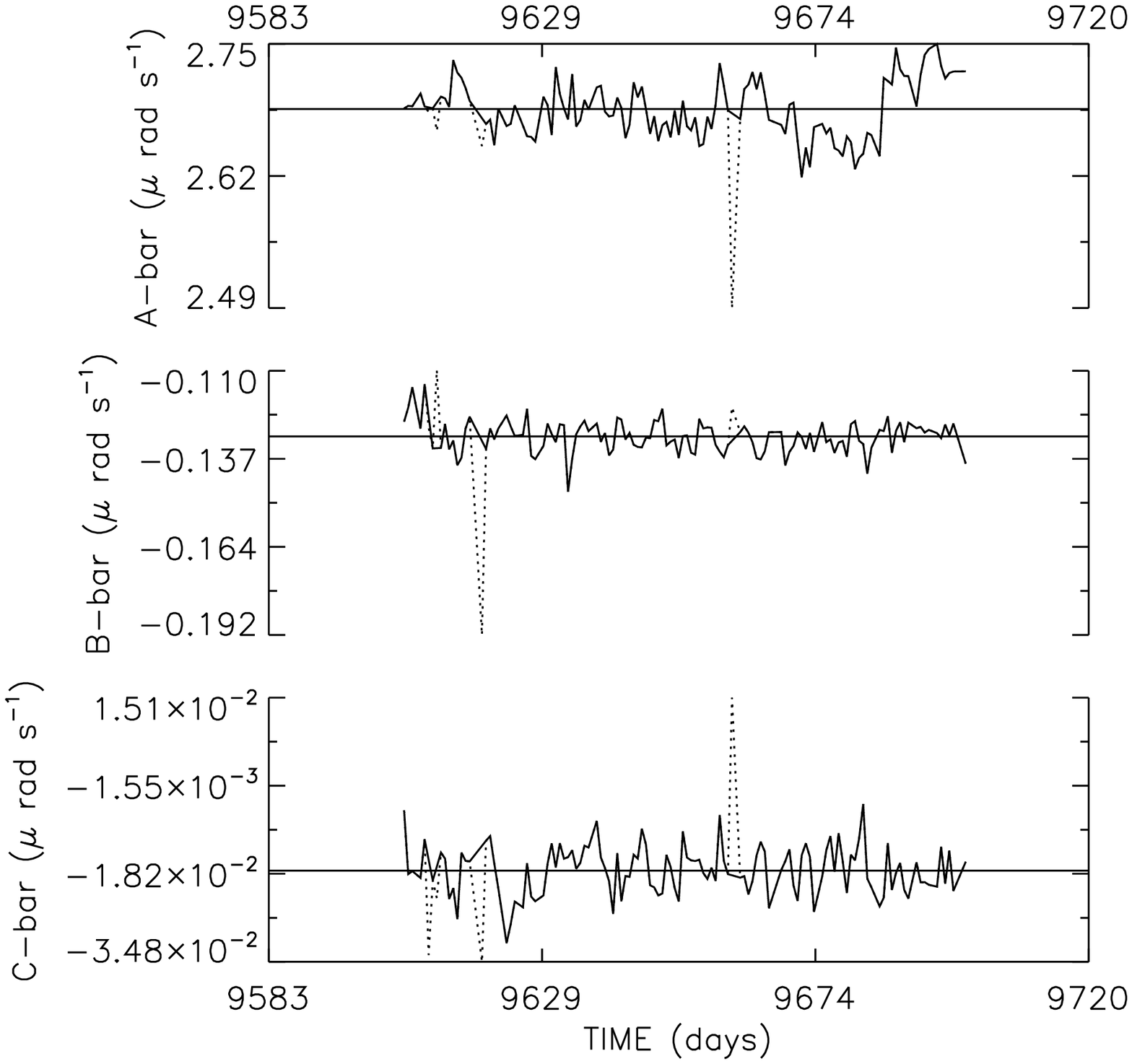}}}
\caption{Variations in the  $\bar A$, $\bar B$ and $\bar C$ ($cf.$ Equation (3)) determined from Mt. Wilson velocity data in
the time (rotation number) intervals 7772\,--\,7934 (top panel-left),
8197\,--\,8297 (top panel-right), 8559\,--\,8726 (middle panel-left),
 8902\,--\,9023 (middle panel-right), 9280\,--\,9376 (bottom panel-left), 
and 9583\,--\,9720 (bottom panel-right). The dotted- and solid-curves
 represent the variations determined from the uncorrected and the 
corrected data, respectively.
The horizontal lines represent the mean values of the corrected data.}
\end{center}
\end{figure}

\begin{figure}
\begin{center}
\subfigure{
\includegraphics[width=5cm]{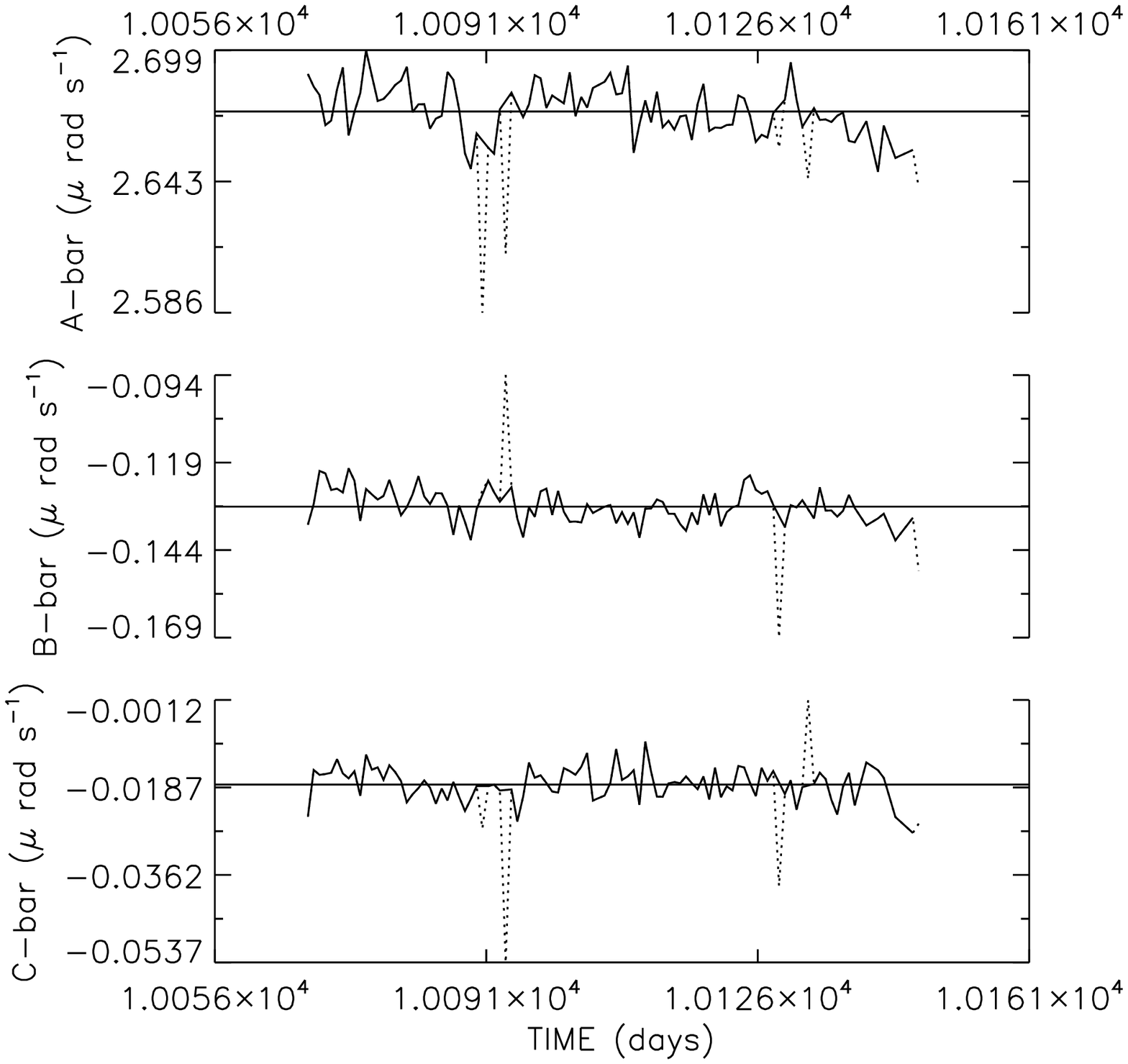}}
\hspace{1.5cm}
\subfigure{
\includegraphics[width=5cm]{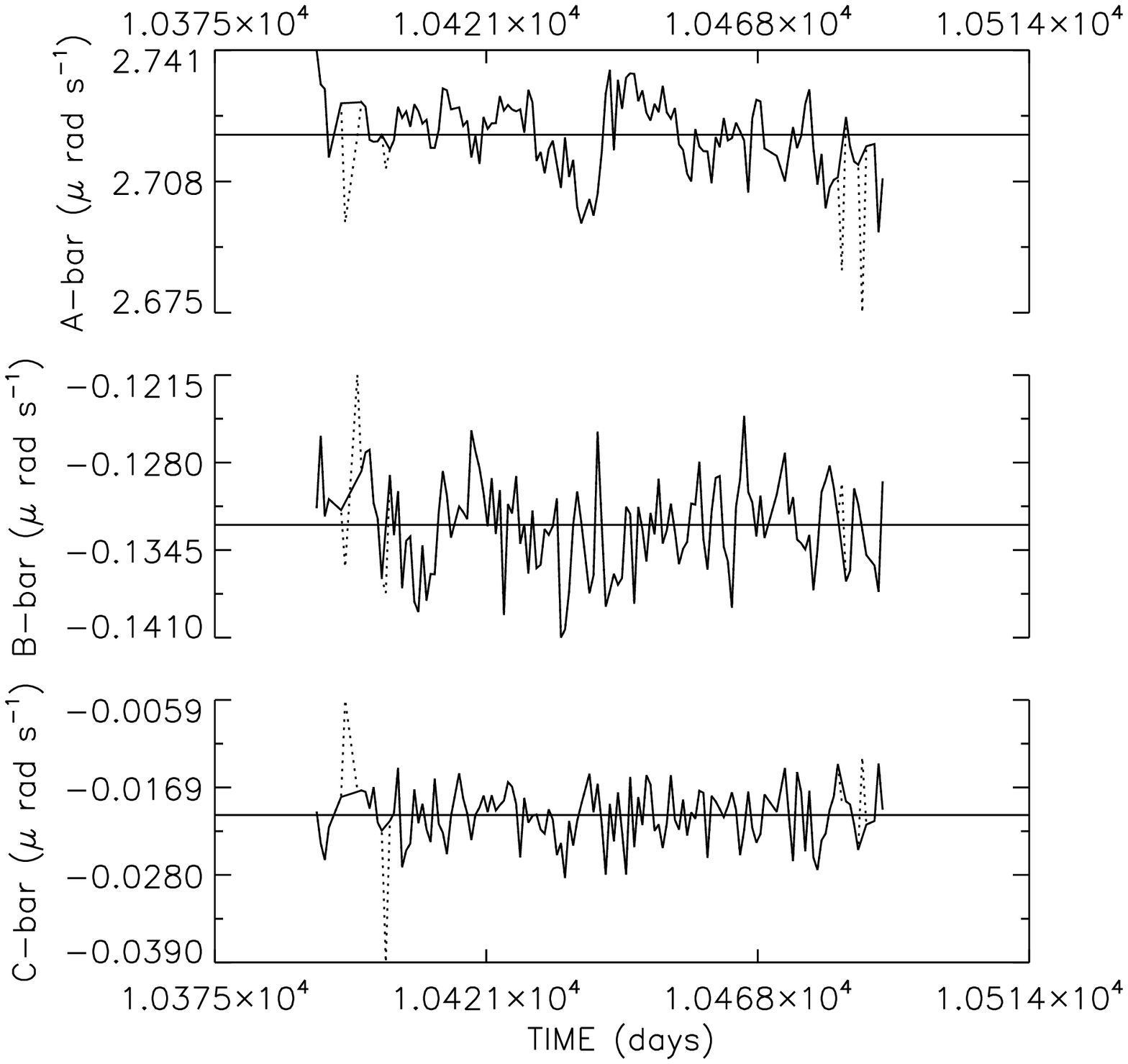}}

\subfigure{
\includegraphics[width=5cm]{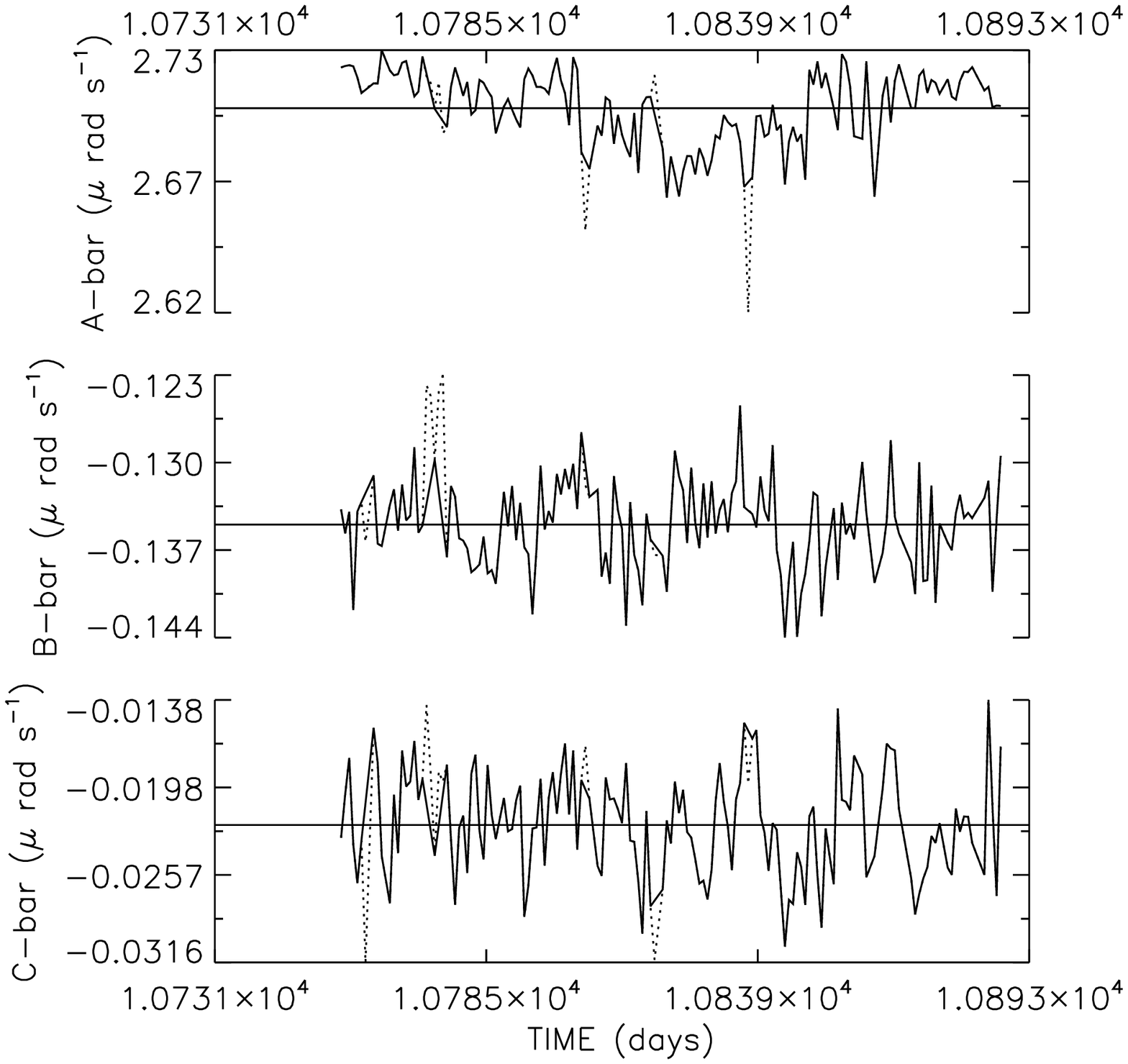}}
\caption{The same as Figure~1, but for the intervals 10056\,--\,10161 
(top panel-left), 10375\,--\,10514 (top-panel-right), and 
10731\,--\,10893 (bottom panel).}
\end{center}
\end{figure}

\begin{figure}
\begin{center}
\includegraphics*[width=\textwidth]{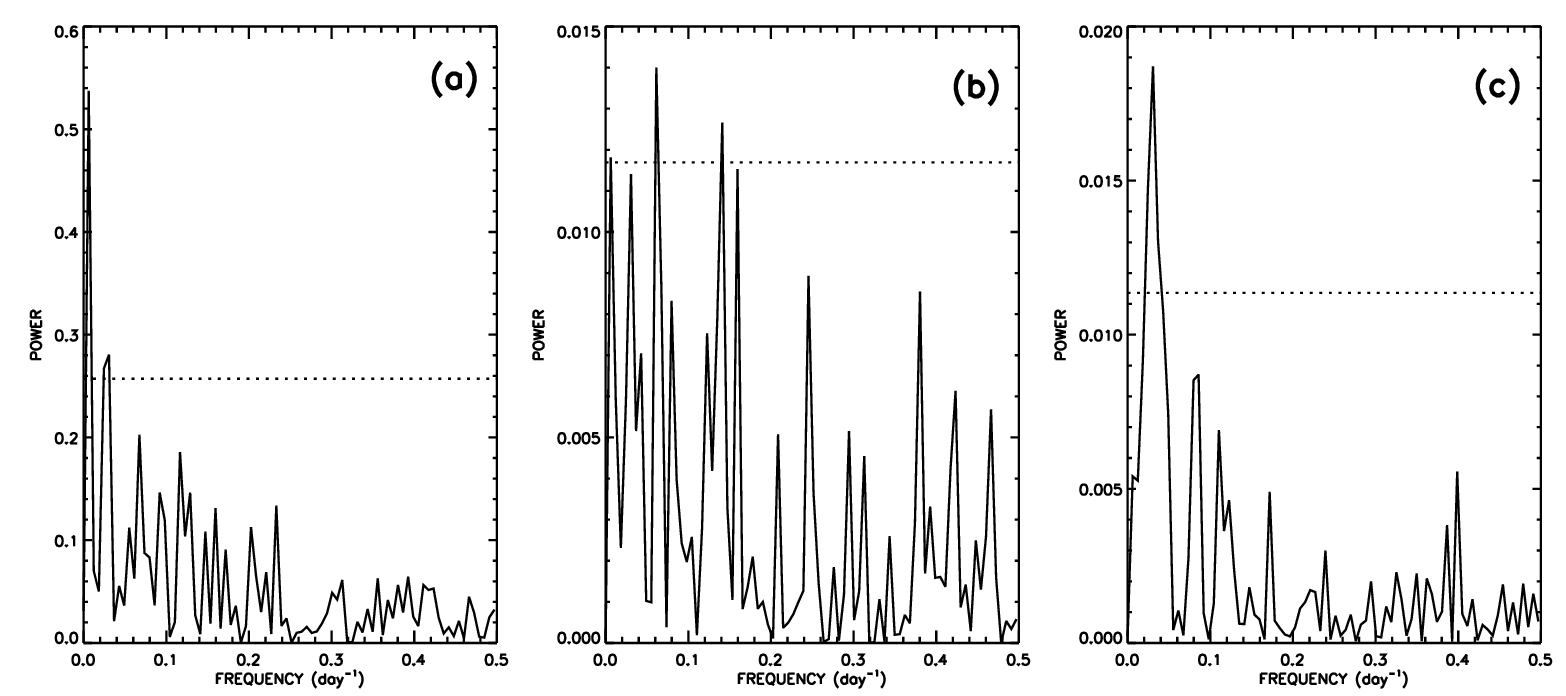}
\includegraphics*[width=\textwidth]{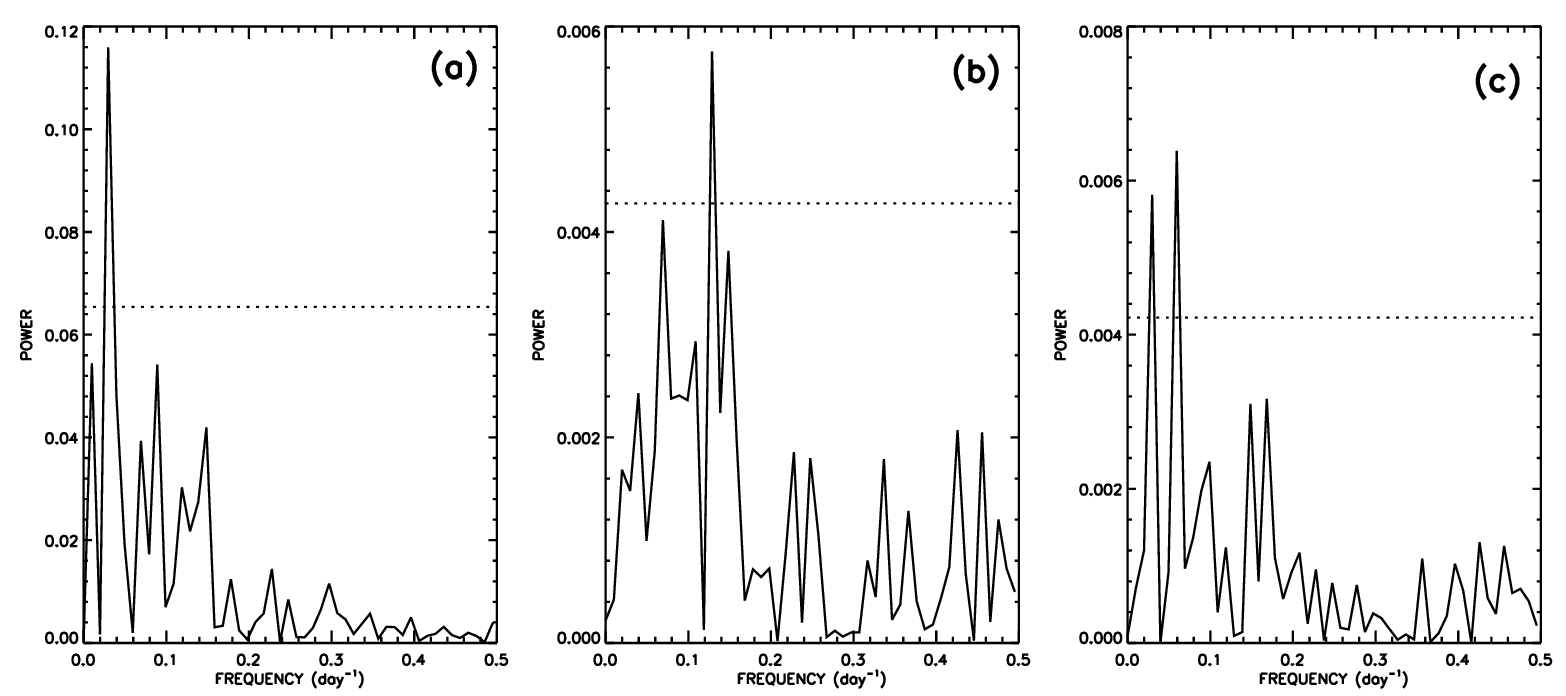}
\includegraphics*[width=\textwidth]{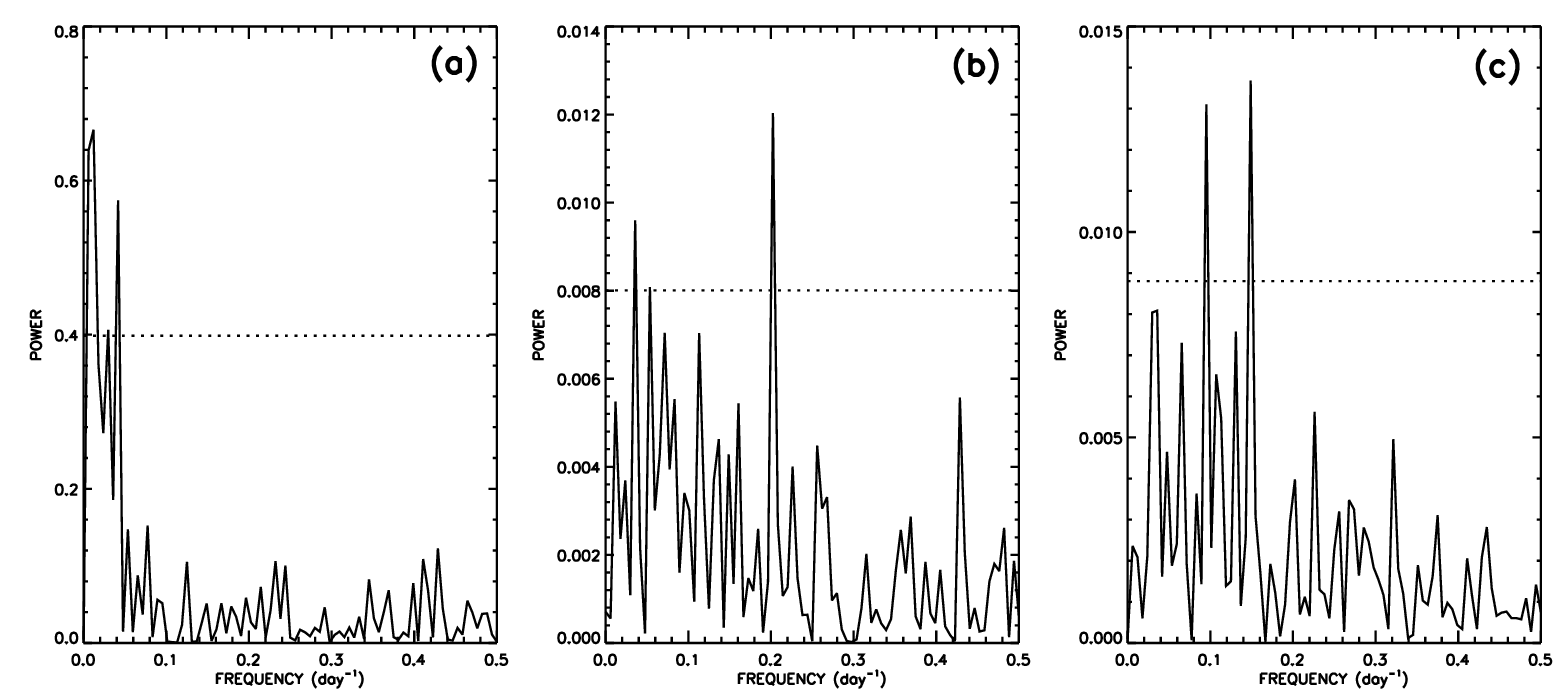}
\caption{FFT power spectra of (a) $\bar A$, (b) $\bar B$ and (c) $\bar C$ 
determined from the Mt. Wilson velocity data in the rotation number 
 intervals
7772\,--\,7934 (top panel),
8197\,--\,8297 (middle panel), and 8559\,--\,8726 (bottom panel).
The dotted horizontal line represents 99\% confidence level.}
\end{center}
\end{figure}

\begin{figure}
\begin{center}
\includegraphics*[width=\textwidth]{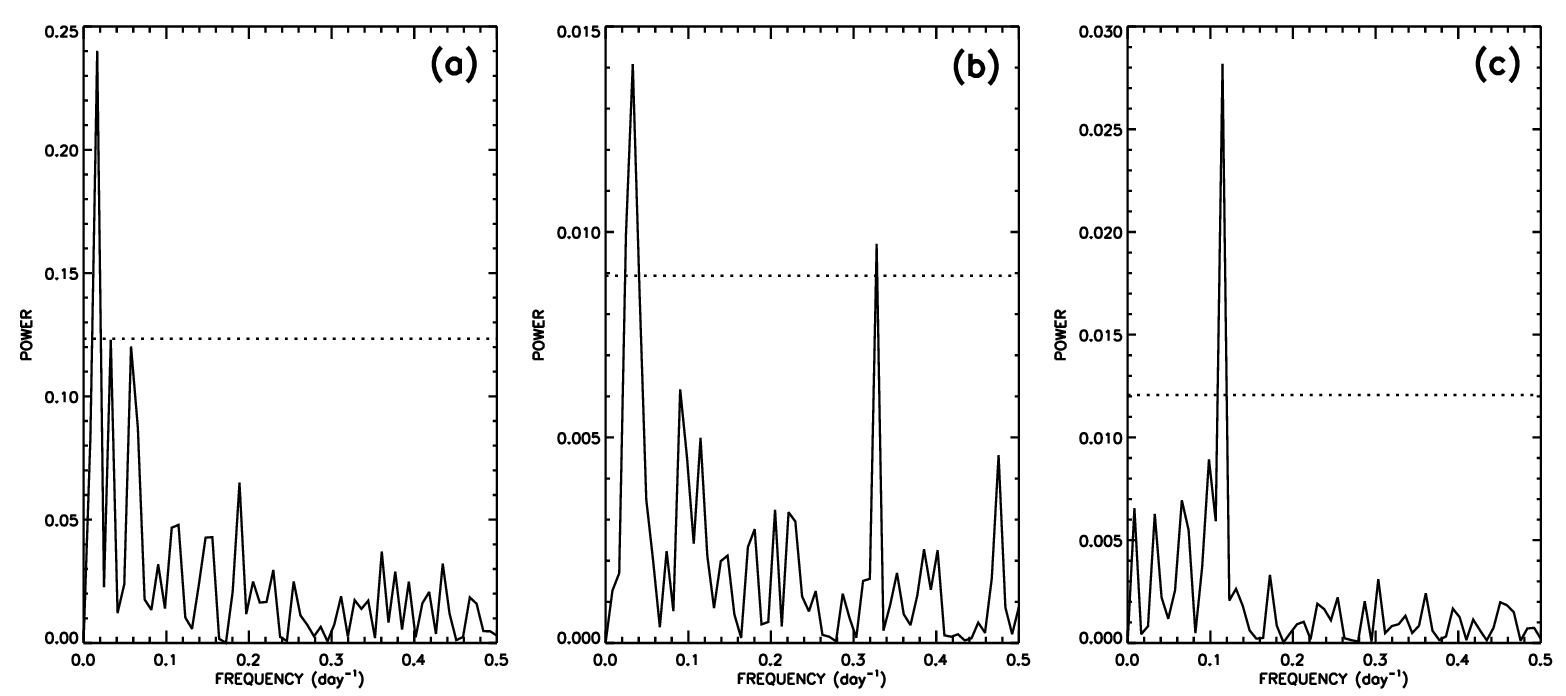}
\includegraphics*[width=\textwidth]{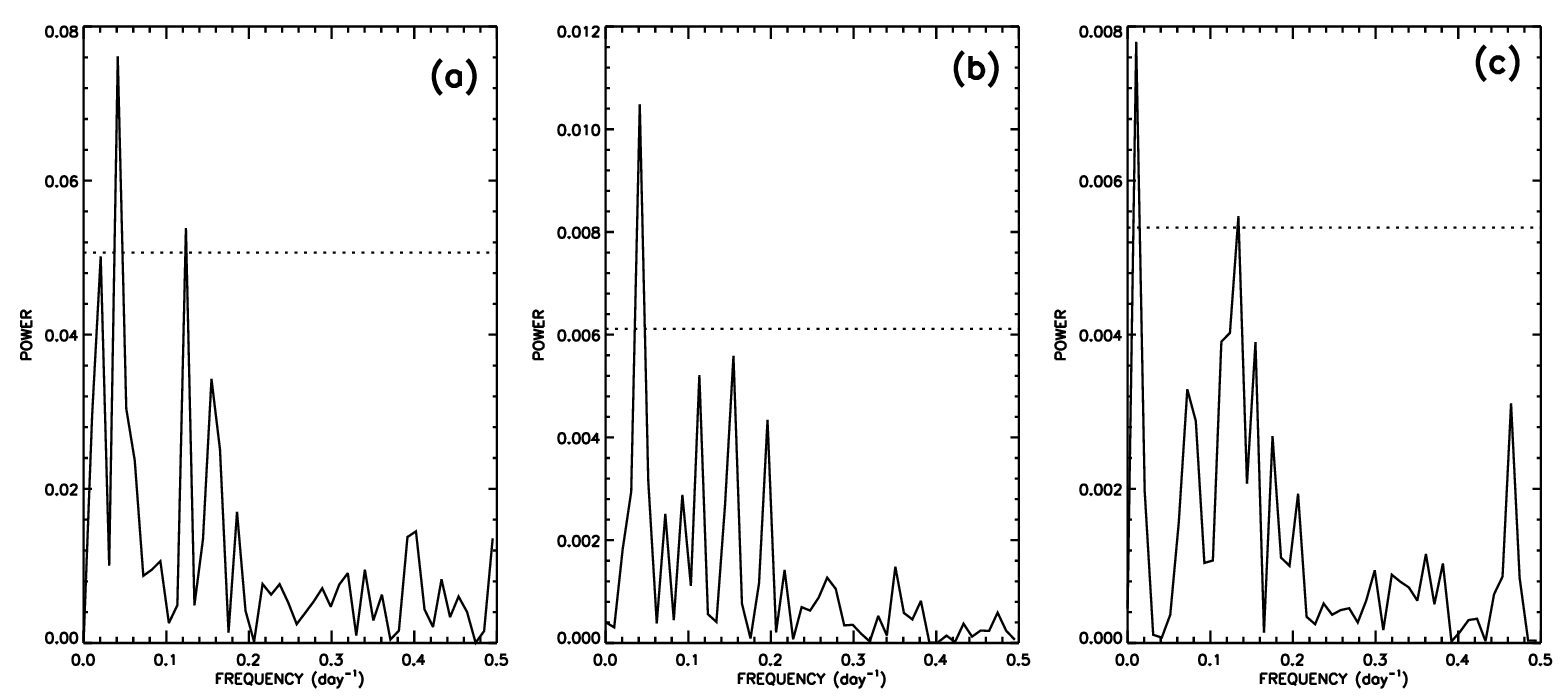}
\includegraphics*[width=\textwidth]{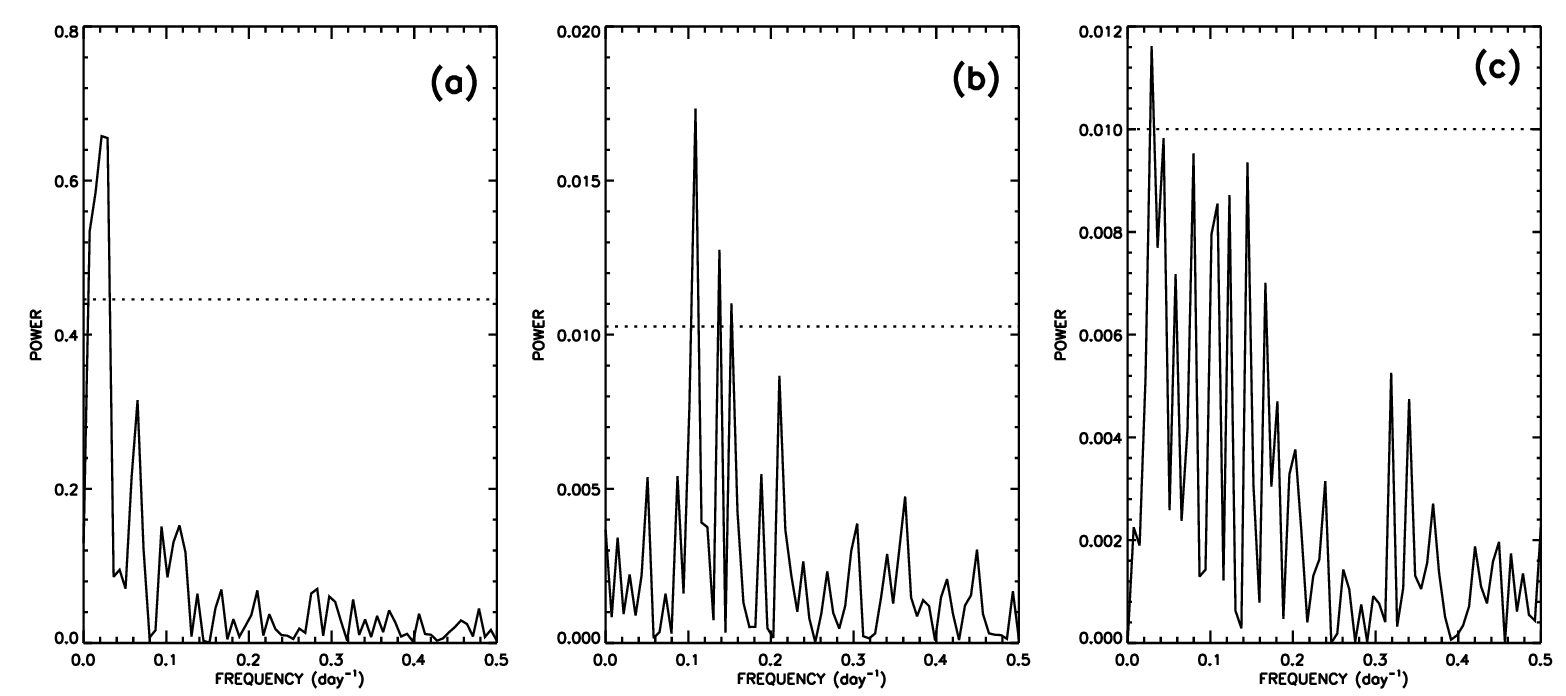}
\caption{The same as Figure~3, but for the intervals
 8902\,--\,9023 (top panel), 9280\,--\,9376 (middle panel),
and 9583\,--\,9720 (bottom panel).}
\end{center}
\end{figure}

\begin{figure}
\begin{center}
\includegraphics*[width=\textwidth]{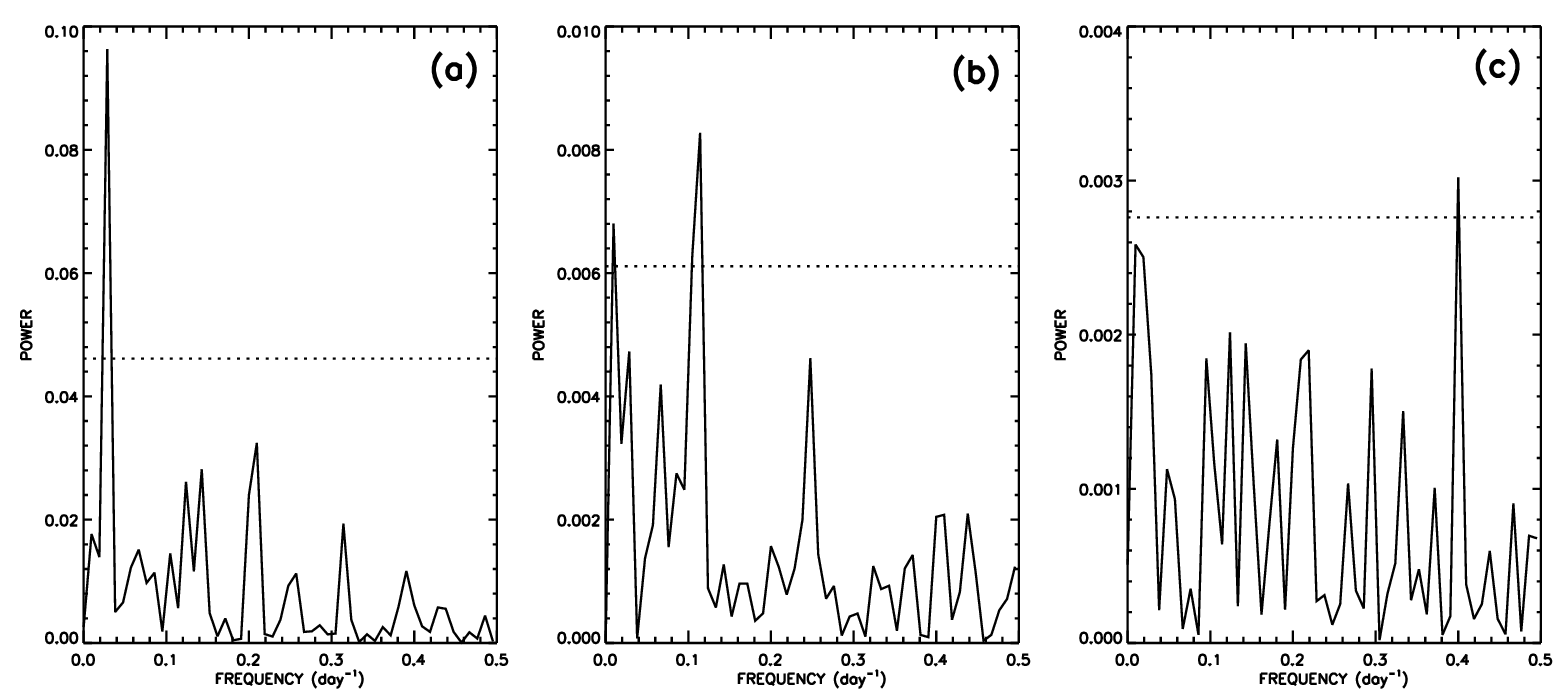}
\includegraphics*[width=\textwidth]{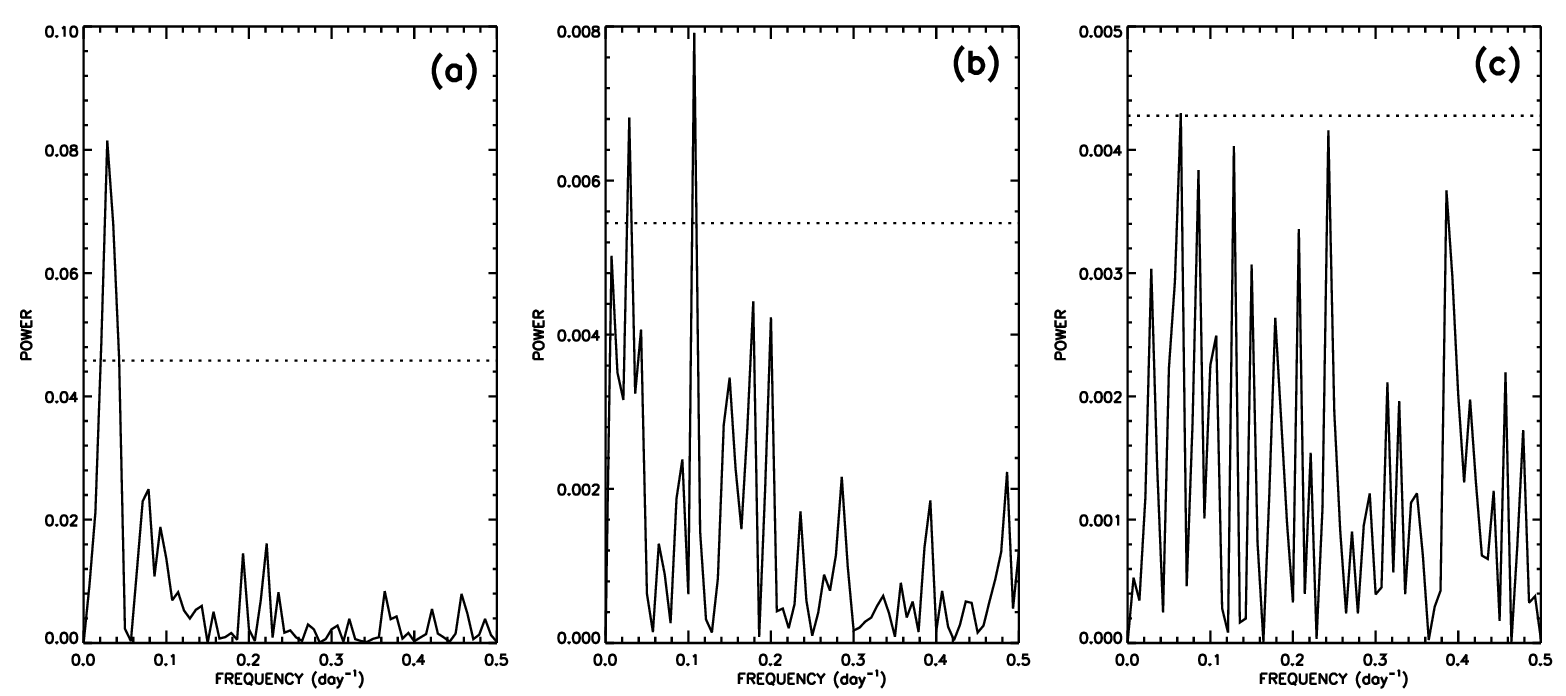}
\includegraphics*[width=\textwidth]{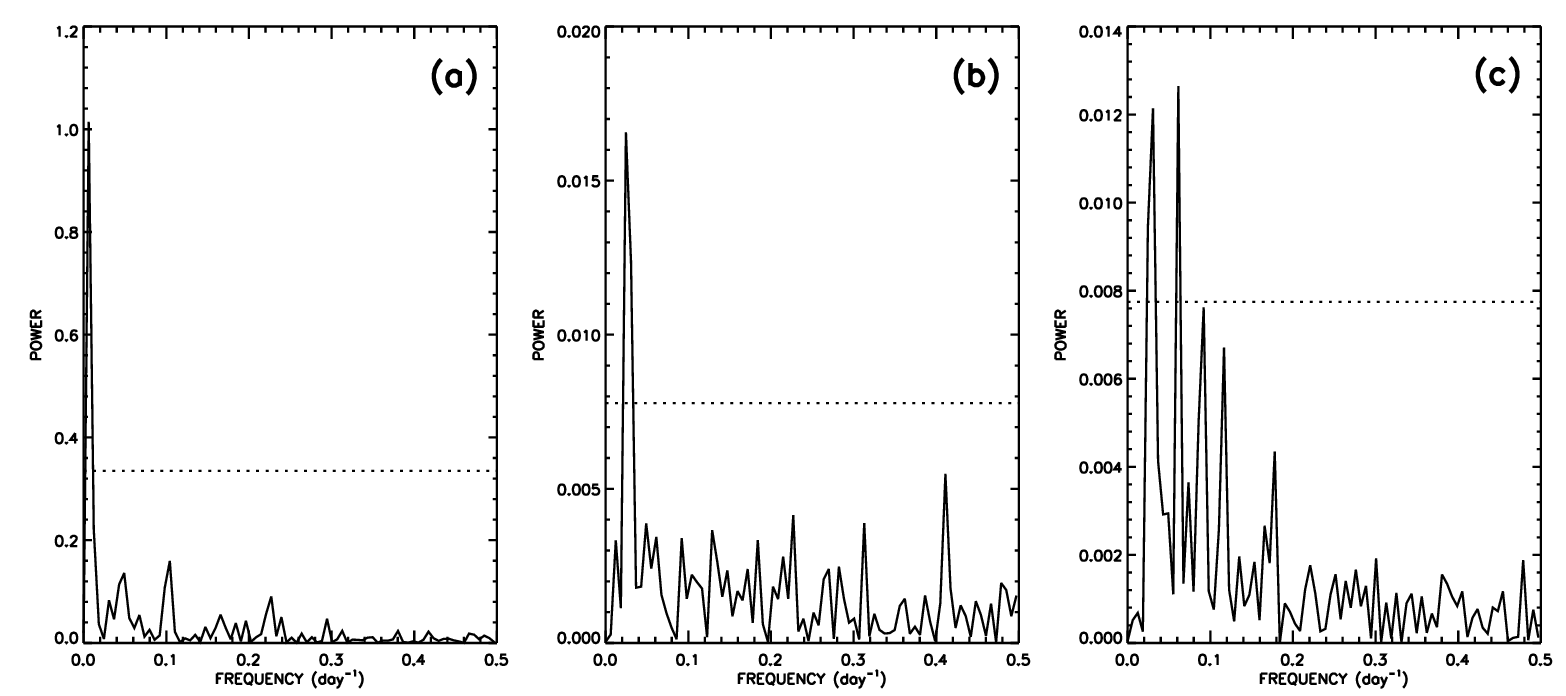}
\caption{The same as Figure~3, but for the intervals 10056\,--\,10161 
(top panel), 10375\,--\,10514 (middle panel), and 
10731\,--\,10893 (bottom panel).}
\end{center}
\end{figure}
\clearpage

\begin{table}[]
\tiny
 \caption{Periodicities in the  variations of $\bar A$, $\bar B$,
and $\bar C$ determined from the daily Mt. Wilson velocity data in different
time intervals mentioned in the first column (the corresponding year and
number of data points are  mentioned within round and square brackets,
respectively).  The levels  of significance (in units of $\sigma$) are also given.}
\smallskip
\begin{tabular} {lcccccccccc} 
     \hline
           \noalign{\smallskip}

 Time  &&  $\bar A$ &&&  $\bar B$ &&&  $\bar C$ \\
 Interval && Period (in days)  & $\sigma$ && Period (in days) & $\sigma$&& Period (in days) & $\sigma$ \\ 
\hline
\noalign{\smallskip}
\\
7772--7934 && $\sim$ 163      &6.2     &&$\sim$ 163   &3.7 &&33-41      &3.5--4.7 \\
(1986), [163]       &&33--41  &2.7--2.9  && $\sim$ 33 $\pm$ 8  &2.4 &&23--27   &2.4--3.0 \\
           &&   &                      &&$\sim$16        & 3.3& & \\
           &&   &                      &&6--7     & 2.5--2.9&& &  \\

\\
8197--8297 &&$\sim$ 101        &2.0     &&$\sim$14    &2.4 &&$\sim$34  &3.8 \\
(1987), [101]  &&$\sim$ 34     &2.0     &&6--7 &2.2--3.84 && $\sim$ 17 &4.3 \\
              &&$\sim$ 11      &2.0 &&&&&&\\     

\\
8559--8726 &&56--84       &2.3--4.6      &&$\sim$ 28 &3.3 &&28-34    &2.9 \\
(1988), [168]  &&$\sim$34       &2.6      &&$\sim$19&2.7   &&$\sim$ 10    &4.3 \\
          && $\sim$24           &3.9      &&$\sim$9 &2.1   &&6--8    &2.0--4.5 \\
          &&                    &         &&$\sim$5 &4.3   &&                &  \\

\\
8902--9023 &&$\sim$ 61       &5.6      &&$\sim$ 30    &4.5 &&$\sim$ 9    &6.7 \\
(1989), [122]  &&$\sim$ 30   &2.6      &&$\sim$24 &2.4 &&  & \\
          &&$\sim$ 17  &2.5            &&$\sim$3   &2.9 &&  & \\

\\
9280--9376 &&$\sim$48     &2.5      &&$\sim$ 24   &4.9    &&$\sim$ 97      &4.1 \\
(1990), [97]  && $\sim$ 24 &4.3      &&$\sim$ 9 &2.1  && $\sim$7  &2.7 \\
           && $\sim$ 8     &2.8        &&$\sim$6 & 2.3               &  &&& \\

\\
9583--9720 &&34--69   &3.5--4.0 &&$\sim$9  &4.9 &&$\sim$ 34      &3.1 \\
(1991), [138]  && &             &&6-7   & 2.8--3.3 &&$\sim$ 12  &2.4 \\
           &&  &                &&$\sim$4 &2.0 &&8--9  &2.0 \\
             &&              &    &&     &             &&$\sim$6  &2.3 \\

\\
10056--10161 &&$\sim$ 35     &5.0 &&$\sim$35  &3.4 &&52--105   &2.2--2.3 \\
(1992), [106]    &&              &    &&8--9   &2.7--3.8 && $\sim$2.5  &2.9 \\

\\
10375--10514 &&$\sim$ 35  &5.0    &&$\sim$ 140  &2.3 && $\sim$15  &2.6 \\
(1993), [140]     && &    &&$\sim$ 35  &3.4          &&$\sim$ 12 &2.2 \\
             &&         &    &&$\sim$ 9   &4.1       && $\sim$ 8  &2.4 \\
             &&           &       &&  &              &&$\sim$ 4  &2.5 \\
             &&           &       &&  &              &&$\sim$ 2.5  &2.0 \\

\\
10731--10893 &&$\sim$ 163 &8.4 &&$\sim$ 32--41  &4.5--6.3 &&$\sim$ 32--41 &3.3--4.4 \\
(1994), [163]     &&           &    && &                       &&$\sim$16    &4.6 \\
     &&           &    && &                       &&$\sim$11    &2.5 \\
     &&           &    && &                       &&$\sim$9    &2.1 \\

\noalign{\smallskip}
\hline
\end{tabular}
\end{table}
\end{document}